\documentclass{article}
\usepackage{graphicx}

\begin{document}

\begin{center}
{\Large\bf A Search for Small-Scale Clumpiness in Orion and W3
High-Mass Star-Forming Regions}

\vspace{4mm}

{\large\bf L.E.~Pirogov$^1$, I.I.~Zinchenko$^1$, \fbox{L.E.B.~Johansson$^2$,}
J.~Yang$^3$}

\vspace{1mm}

{\it
$^1$ -- Institute of Applied Physics, Russian Academy of Sciences \\
$^2$ -- Onsala Space Observatory, Sweden \\
$^3$ -- Purple Mountain Observatory, China
}

\end{center}

\begin{abstract}
Observations of distinct positions in Orion and W3 revealed
ripples on the HCN(1--0), HCO$^+$(1--0) and CO(1--0) line profiles
which can be result of emission of large number
of unresolved thermal clumps in the beam that move with random velocities.
The total number of such clumps are $\sim (0.4-4)\cdot 10^5$ for the areas
with linear sizes $\sim 0.1-0.5$~pc.
\end{abstract}

\section{Introduction}

There are many evidences that dense molecular cloud cores,
where high-mass stars and stellar clusters are born,
are clumpy on different spatial scales down to the scales unresolved
by telescope beams.
In particular, this follows from nearly constant volume densities in clouds
with strong column density variations \cite{Ber96} or from
detection of C~I emission over large areas correlated with
molecular maps \cite{WP91}.
An important evidence for small-scale clumpiness in high-mass star-forming
(HMSF) regions is provided by observations
of low relative intensity of the $F=1-1$ HCN(1--0) hyperfine component
compared to the optically thin case, which is connected with overlaps
of thermal profiles of hyperfine components in higher HCN transitions.
This phenomenon together with the fact that the observed line widths
are highly suprathermal can be explained in the model with randomly moving
thermal unresolved clumps \cite{Pir99}.

Crude estimates for physical parameters of unresolved clumps
can be obtained from simple analytical model \cite{Mar84,Tau96}.
Assuming that clumps are identical
with a small volume filling factor while velocity dispersion
of their relative motions ($\sigma$) is much higher
than inner velocity dispersion ($v_0$),
we should expect an appearance of ripples in line profiles
due to fluctuations in the number of clumps on the
line of sight at various velocities.
An extent of such ripples could be determined by
the standard deviation for radiation temperature fluctuations
in some range near the line center ($\Delta T_{\rm R}$).
In particular, total number of clumps in the telescope beam ($N_{\rm tot}$)
can be derived from the following expression \cite{Tau96}:

$$
\frac{\Delta T_{\rm R}}{T_{\rm R}}=
\frac{\tau}
{(e^{\tau}-1)\sqrt{N_{\rm tot} \frac{v_0}{\sigma}}} \hspace{2mm},
$$

\noindent{where $\tau$ is line optical depth, that can be obtained
from width comparison of the optically thick and thin Gaussian lines.}
Using the observed values of $\Delta T_{\rm R}$, peak line intensities
($T_{\rm R}$) and $\tau$ for two lines with different optical depths
and knowing kinetic temperature it is possible to estimate
the number of thermal clumps in the telescope beam.

Previously we used this approach
in the analysis of the CS(2--1) and HCN(1--0) and their rare isotopes
line profiles observed towards dense cores associated
with S140, S199 and S235 HMSF regions \cite{PZ08}.
The total number of clumps in the beam have been derived.
Besides, clump densities, sizes and volume filling factors
have been obtained from detailed model calculations.
Here we present preliminary results of the line profile analysis
towards selected positions in Orion and W3.

\section{Small-scale clumpiness in Orion and W3}

It is known that inhomogeneous clumpy structure is a feature
of photon-dominated regions (PDR)
that are usually located on the periphery of molecular clouds.
One of the most well-known objects of that class is Orion Bar region.
The HCN(1--0) spectra observed in this region show anomalies
typical for massive dense cores \cite{Owl00}.
Thus, one could also expect an existence of unresolved thermal clumps
in this region.

In order to search and compare parameters of small-scale clumpy structure
in dense core and PDR we selected position of maximum HCN(1--0) intensity
in the Orion~Bar region.
For comparison the Orion~KL position has been taken.
Also, we selected two positions in W3 HMSF complex.
The central part of this complex is known to has a rarefied structure,
similar to clumpy PDR \cite{Kram04}.
We selected position of maximum HCN(1--0) intensity near W3~IRS5
and W3(OH).

Long-time integration in the HCN(1--0) and HCO$^+$(1--0) lines
that trace of high density gas ($\sim 10^5-10^6$~cm$^{-3}$)
have been done at 20-m radiotelescope of Onsala Space Observatory.
Rare isotopic lines, H$^{13}$CN(1--0) and H$^{13}$CO$^+$(1--0),
have been used to determine optical depth of main lines.
In order to determine parameters of lower density gas
($\sim 10^3-10^4$~cm$^{-3}$) that can trace interclump gas,
the same positions have been observed in the CO(1--0) line
at the 13,7-m telescope of Purple Mountain Observatory.

The list of the sources with coordinates, distances and linear resolutions
at the frequencies of observations are given in Table~1.
Integration times for the HCN, HCO$^+$ and CO lines vary 
from 2 hours to 11 hours.

\begin{table}[hbtp]
\begin{center}
\caption{Source list}
\smallskip
\begin{tabular}{|c|c|c|c|c|}\hline\noalign{\smallskip}
Source          & $\alpha$(2000) & $\delta$(2000) & D      & Resolution\\
                  &${\rm (^h)\  (^m)\  (^s)\ }$
                                   &$(^o)$  $(^{\prime})$  $(^{\prime\prime}$)
                                                    & (kpc)  & (pc)\\
\noalign{\smallskip}\hline\noalign{\smallskip}
Orion KL           &  05 35 14.5 &--05 22 27  & 0.45 & 0.09--0.11 \\
Orion Bar          &  05 35 20.1 &--05 26 07  & 0.45 & 0.09--0.11 \\
W3 IRS5            &  02 25 40.7 &  62 05 52  & 2.3  & 0.45--0.56 \\
W3 (OH)            &  02 27 04.7 &  61 52 26  & 2.3  & 0.45--0.56 \\
\noalign{\smallskip}\hline\noalign{\smallskip}
\end{tabular}

\label{table:list}
\end{center}
\end{table}

The observed line profiles towards Orion~KL and Orion~Bar are given 
in Fig.1.
Below each profile residual fluctuations obtained after FFT filtering 
of the main profile component are given. 
The rejection level is equal to 0.8~(km/s)$^{-1}$
(see \cite{PZ08} for detailed description of this method).

\begin{figure}[hbtp]
 \centering \includegraphics[width=12cm]{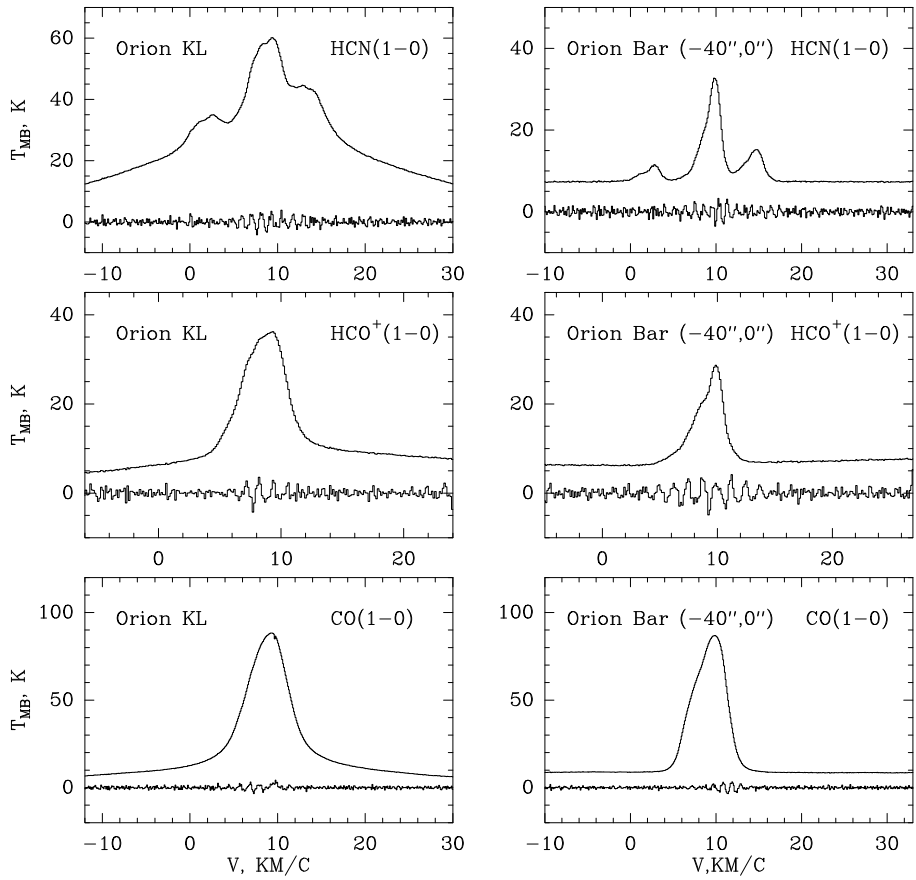}
\caption{The observed line profiles towards Orion~KL and Orion~Bar 
(left and right panels, respectively).
Residual fluctuations multiplied by a factor of ten 
are shown below each profile.
}
\end{figure}

Both for PDR and dense core positions in Orion dispersions
of residuals within the line range for the HCN, HCO$^+$ and CO lines
are significantly higher than dispersion of noise.
We also detected significant residuals for the HCO$^+$ line
in W3~IRS5 and for the CO line in W3(OH).
Unfortunatelly, the widths of Gaussian HCN and HCO$^+$ profiles in Orion Bar
and of Gaussian HCO$^+$ profile in Orion~KL are close to those of rare isotopes
preventing to derive optical depth and total number of clumps.
Yet, the H$^{13}$CN profile towards Orion KL (ridge component) consist of
two triplets we fitted it with the single one to compare with results of HCN 
fitting.
Therefore, one should treat the results for Orion KL with caution.

Peak line intensities,
standard deviations of residual fluctuations
within line range, and optical depths are given in Table~2
for Orion~KL, W3~IRS5 and W3(OH).
The total number of clumps in the beam is given in the last column.
For W3(OH) we calculated $\tau$ and $N_{\rm tot}$ using CO
and C$^{18}$O \cite{NRAO} data.
The subsequent data analysis will include 
detailed model calculations which should help 
to derive physical parameters of clumps.

\begin{table}[hbtp]
\begin{center}
\caption{Line parameters and total number of clumps in the beam}
\smallskip
\begin{tabular}{|c|c|c|c|c|c|}\hline\noalign{\smallskip}
Source & Line & $T_{\rm R}(K)$ & $\Delta T_{\rm R}(K)$ & $\tau$ & $N_{\rm tot}$ \\
\noalign{\smallskip}\hline\noalign{\smallskip}
Orion KL (ridge) & HCN(1--0)        & $\sim 31$         & 0.21  & $\sim 1.5$    &$\sim 40000$ \\
\noalign{\smallskip}\hline\noalign{\smallskip}
W3 IRS5   & HCO$^+$(1--0) & $\sim 15$ & 0.04         & $\sim 1.2$    &$\sim 400000$ \\
\noalign{\smallskip}\hline\noalign{\smallskip}
W3(OH)    & CO(1--0)      & $\sim 25$ & 0.08         & $\sim 2.6$    &$\sim 90000$ \\
\noalign{\smallskip}\hline\noalign{\smallskip}
\end{tabular}

\label{table:ntot}
\end{center}
\end{table}

\vspace{1mm}

{\small The work is supported by the Russian Foundation for Basic Research
(projects 06-02-16317 and 08-02-00628) and the Basic
Research Program of the Division of Physical Sciences of the Russian Academy
of Sciences on ``Extended Objects in the Universe".}

{}


\begin{thebibliography}{99}

\bibitem{Ber96} Bergin E.A., Snell R.L., and Goldsmith P.F., Astrophys.~J.
460, 343 (1996).

\bibitem{WP91} White G.J., Padman R., Nature 354, 511 (1991).

\bibitem{Pir99} Pirogov L., Astron.~Astrophys., 348, 600 (1999).

\bibitem{Mar84} Martin H.M., Sanders D.B., and Hills R.E., Mon.~Not.~R.~A.~S.
208, 35 (1984).

\bibitem{Tau96} Tauber J.A., Astron.~Astrophys. 315, 591 (1996).

\bibitem{PZ08} Pirogov L.E., Zinchenko I.I. Astron.~Zh. 85, 1072 (2008);
Astron.~Reports 52, 963 (2008)

\bibitem{Owl00} Owl R.C.Y., Meixner M.M., Wolfire M., Tielens A.G.G.M.,
Tauber J., Astrophys.~J. 540, 886 (2000).

\bibitem{Kram04} Kramer C., Jakob H., Schneider N., Br\"ull M., Stutzki J.
Astron.~Astrophys. 424, 887 (2004).


\bibitem{NRAO} http://www.cv.nrao.edu/\~\,jmangum/12meter/stanspec.html

\end{thebibliography}
\end{document}